# Effects of loss factors on zero permeability and zero permittivity gaps in 1D photonic crystal containing DNG materials

Alireza Aghajamali [1] *, Tannaz Alamfard [1], and Mahmood Barati [2]

1- Department of Physics, Marvdasht Branch, Islamic Azad University, Marvdasht, Iran
2- Department of Physics, Science and Research Branch, Islamic Azad University, Fars, Iran

**Abstract**

The effects of electric and magnetic loss factors on zero-$\mu$ and zero-$\varepsilon$ gaps in a one-dimensional lossy photonic crystal composed of double-negative and double-positive materials are theoretically investigated by employing the characteristic matrix method. This study contradicts the previous reports as it indicates that by applying the inevitable factors of electric and magnetic losses to the double-negative material, the zero-$\mu$ and zero-$\varepsilon$ gaps appear simultaneously in the transmission spectrum, being independent of the incidence angle and polarizations. Moreover, the results show that these gaps appear not only for an oblique incidence but also in the case of normal incidence, and their appearance at the normal incidence is directly related to the magnetic and electric loss factors. Besides, the results indicate that as the loss factors and angle of incidence increase, the width of both gaps also increases.

## 1. Introduction

Photonic crystals (PCs) are artificial dielectric or metallic structures in which the refractive index changes periodically in space. Being a relatively new kind of material, PCs were theoretically introduced by Yablonovitch in 1987, and John introduced them experimentally in the same year [1,2]. PCs' unique electromagnetic characteristics along with their scientific and engineering applications have attracted the attention of researchers over the past two decades. The propagation of electromagnetic waves can be affected by this type of periodic structure just as the movement of electrons is affected by the periodic potential of semiconductor crystal, as we define the allowed and forbidden electronic energy bands. The frequencies of photons determine whether or not photons can propagate through PC structures. Modes refer to the frequencies that let the photons travel through the structure, and a group of such modes is referred to as band. Photonic band gaps (PBGs), then, represent the disallowed bands of frequency [3,4]. These PBGs are also called the Bragg gaps which are formed as a result of Bragg scattering in a periodically of dielectric structure through which no electromagnetic wave can propagate. In a simple one-dimensional (1D) periodic multilayer structure or a 1D photonic crystal, this is the properties of PBGs that cause PCs to have interesting applications in the fields of photonics and optical engineering, as in photonic devices, optical filters, resonance cavities, laser applications, high reflecting omnidirectional mirrors, and the optoelectronic circuits [5-8].

---

* Corresponding author. Tel: +98 917 703 6389.
  E-mail address: alireza.aghajamali@fsriau.ac.ir (A. Aghajamali)

Velelago [9] was the one who predicted the existence of materials with negative refractive index and simultaneous negative permittivity and permeability. These materials, which are now called double-negative (DNG) materials, have received extensive attention for their very unusual electromagnetic properties. Negative refractive index materials, simplified as negative index materials (NIMs), are now also known as left-handed materials or metamaterials. Since 2000, when Smith et al. [10] experimentally realized metamaterials at microwave frequencies, their unusual electromagnetic frequencies have gained worldwide attention.

As regards DNG material, it has been demonstrated that the general entropy conditions for the constitutive items make this material dispersive [11]. Inherent metal loss is found to be inevitable in DNG materials, as they are artificially made of metal thin-wire (TW) and metal split-ring resonator (SRR) constructions [10,11]. As a result, DNG materials are not only dispersive but also lossy [12]. It has been indicated that DNG materials have complex value of electric permittivity and magnetic permeability. The permittivity and the permeability have real parts which are simultaneously negative. Loss factors are expressed by the imaginary parts of the permittivity and the permeability.

During the past decades, the possibility of using metamaterials in the production of photonic crystals, which is called metamaterial photonic crystals (MetaPCs), has given rise to a new research area whose interesting results have so far been reported by various researchers. Among the papers written and published on MetaPCs properties, we direct the audience's attention to an interesting report by Li et al. [13] in 2003, in which the authors reported the appearance of an additional gap called zero-$\bar{n}$ gap in the transmission spectra of a 1D MetaPC composed of NIM and positive index material dielectric layers. The zero-$\bar{n}$ gap is independent of periodicity as compared to the conventional Bragg gap in which periodicity has an important role due to the fact that it is formed by the destructive interferences of electromagnetic waves in the PC. In recent years, the details of properties and characteristics of the zero-$\bar{n}$ gap of 1D MetaPC structure consisting of DNG and double-positive (DPS) materials have been investigated. Some examples of these investigated properties include the bandwidth, the depth, and the central frequency of the gap [13-19]. In 2007, Depine et al. [20] and Singh et al. [21], who published their reports approximately at the same time, demonstrated that in 1D MetaPC, in addition to the zero-$\bar{n}$ and conventional Bragg gaps, two new gaps appear in the transmission spectra for both TE and TM polarized waves. In the TE mode, the appearance of the new gap, which is called the zero-$\mu$ gap, occurs at the frequency where the permeability of the DNG material disappears, but in the TM mode, another gap called the zero-$\varepsilon$ gap appears at the frequency wherein the permittivity of the DNG material reaches zero. Previous report, thus, have noted that the zero-$\mu$ and the zero-$\varepsilon$ gaps only appear at oblique incidence angles, and therefore, are absent at normal incidence angles [20,21].

The main purpose of this work is to conduct a theoretical investigation of the effects of electric and magnetic loss factors on the zero-$\mu$ and zero-$\varepsilon$ gaps in a 1D lossy MetaPC composed of DNG and DPS materials. Our study focuses on the behavior of the zero-$\mu$ and zero-$\varepsilon$ gaps by applying the $\gamma_e$ and $\gamma_m$ to the DNG material for both normal and oblique incidence cases. The outline of this paper is as follows: Section 2 deals with the geometric MetaPC structure, the characteristic matrix method and its formulation, and also the permittivity and permeability of DNG material, section 3 presents the numerical results and discusses the normal and oblique incidence cases, and section 4 puts forward the conclusion of the study.

## 2. MetaPC structure and characteristic matrix method

This study is conducted on a 1D MetaPC located in air, which consists of alternative layers of DPS materials and dispersive and dissipative DNG materials. The investigated 1D MetaPC has the periodic structure $(AB)^N$, where layer A is assumed to be consisted of DNG material and layer B of DPS material. The number of the lattice period is referred to as N, and thickness, permittivity, and permeability of the layers are respectively referred to as $d_i$, $\varepsilon_i$, and $\mu_i$ ($i = A, B$).

In performing calculations, we have used the characteristic matrix method [22] that is considered the most effective technique to analyze the transmission properties of 1D PCs. For the TE wave at the incidence angle $\theta_0$, from vacuum to a 1D PC structure, the characteristic matrix $M[d]$ is calculated through the following equation [22]:

$$M[d] = \prod_{i=1,2} \begin{bmatrix} \cos\gamma_i & \frac{-i}{p_i}\sin\gamma_i \\ -ip_i\sin\gamma_i & \cos\gamma_i \end{bmatrix}, \qquad (1)$$

where $\gamma_i = (\omega/c)n_i d_i \cos\theta_i$, $c$ is the speed of light in vacuum, $\theta_i$ is the ray angle inside layer $i$ with refractive index $n_i$, $p_i = \sqrt{\varepsilon_i/\mu_i}\cos\theta_i$, and $\cos\theta_i = \sqrt{1-(n_0^2\sin^2\theta_0/n_i^2)}$, in which $n_0$ is the refractive index of the environment wherein the incidence wave tends to enter the structure. The refractive index is given as $n_i = \pm\sqrt{\varepsilon_i \mu_i}$ [18,23], for the DPS material, as usual, we choose the positive sign while the negative sign is considered in the DNG material. The characteristic matrix for N period structure is, therefore, $[M(d)]^N$. The transmission coefficient of the multilayer is calculated by:

$$t = \frac{2p_0}{(m_{11} + m_{12} p_s) p_0 + (m_{21} + m_{22} p_s)}. \qquad (2)$$

In this equation, $m_{ij}$ ($i, j = 1,2$) are the matrix elements of $[M(d)]^N$, $p_0 = n_0 \cos\theta_0$, and $p_s = n_s \cos\theta_s$, with $n_s$ being the refractive index of the environment where the wave leaves the crystal with the angle $\theta_s$. The transmissivity of the multilayer is given by $T = (p_s/p_0)|t|^2$. The transmissivity of the multilayer for TM wave can be obtained by using the previous expressions, with $p_i = \sqrt{\mu_i/\varepsilon_i}\cos\theta_i$, $p_0 = \cos\theta_0/n_0$, and $p_s = \cos\theta_s/n_s$.

The permittivity and permeability of a material determine its electromagnetic properties. The following equations give the complex permittivity and permeability of layer A with negative refracting index in the microwave region [13,14]:

$$\varepsilon_A(f) = 1 + \frac{5^2}{0.9^2 - f^2 - if\gamma_e} + \frac{10^2}{11.5^2 - f^2 - if\gamma_e}, \qquad (3)$$

$$\mu_A(f) = 1 + \frac{3^2}{0.902^2 - f^2 - if\gamma_m}. \qquad (4)$$

In equations (3) and (4), $f$ is the frequency measured in GHz, and $\gamma_e$ and $\gamma_m$ are respectively the electric and magnetic damping frequencies (also called the electric and magnetic loss factors) which are given in GHz again. Fig. 1 demonstrates various details of the real parts of the permittivity and permeability of layer A, $\varepsilon'_A$ and $\mu'_A$, versus frequency for two different electric and magnetic loss factors.

As shown and mentioned in our previous reports [14,24-27] and observed from the figure, there exists three different regions. In the first region, $f < 3.13$ GHz, and $\varepsilon'_A$ and $\mu'_A$ are simultaneously negative (DNG material). For $3.13 < f < 3.78$ GHz, in the second region, $\varepsilon'_A < 0$ but $\mu'_A > 0$ (single-negative material extending to the epsilon-negative material). Finally, in the third region where $f > 3.78$ GHz, both $\varepsilon'_A$ and $\mu'_A$ are positive (DPS material). It must be noted that the numbers 5, 10, 0.9, 11.5, 3, and 0.902 shown in equations (3) and (4), as well as $f$, $\gamma_e$, and $\gamma_m$, are all in the unit of GHz.

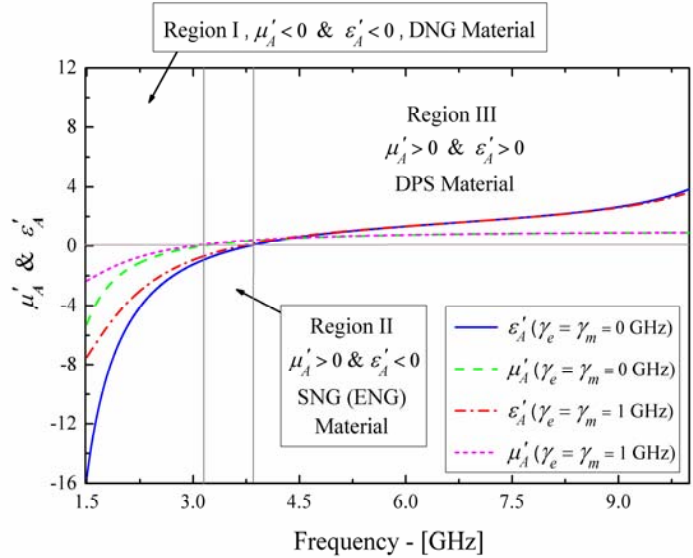

Fig. 1. Real parts of permittivity and permeability of layer A, $\varepsilon'_A$ and $\mu'_A$, versus frequency for $\gamma_e = \gamma_m = 0$ and 1 GHz.

## 3. Numerical results and discussion

We can calculate the transmission spectrum of the present lossy MetaPC structure based on the theoretical model described in the previous section. Equations (3) and (4) calculate the permittivity ($\varepsilon_A$) and the permeability ($\mu_A$) of layer A in the 1D lossy MetaPC consisting of DNG material (in layer A) and DPS material (in layer B). Layer B is assumed to be a vacuum layer with $\varepsilon_B = \mu_B = 1$ ($n_B = 1$). The thickness of layers A and B are respectively chosen as $d_A = 6$ mm and $d_B = 12$ mm, and also $N = 16$ is selected as the total number of lattice period [14].

In Figs. 2(a) and 2(b), we have plotted the frequency-dependence transmission in TE and TM polarized waves for three different incidence angles $\theta_0 = 0°, 30°$, and $60°$, when the loss factors are not considered, i.e., $\gamma_e = \gamma_m = 0$. It can be observed in these figures that, as mentioned in [20,21], as the angle of incidence increases, in addition to the Bragg and zero-$\bar{n}$ gaps, a new gap appears between these two gaps in the transmission spectra for both TE and TM waves. In the TE mode, the new gap emerges at the frequency in which the real part of the permeability ($\mu'$) of DNG material vanishes and changes its sign. Thus, this new gap is called zero-$\mu$ gap. In the TM mode, this new gap appears at the frequency where the real part of permittivity, or $\varepsilon'$, of DNG material goes to zero, so it is known as zero-$\varepsilon$ gap. Moreover, as mentioned in the previous researches [20,21], these band gaps arise only for the oblique incidence and they do not exist at the normal incidence case.

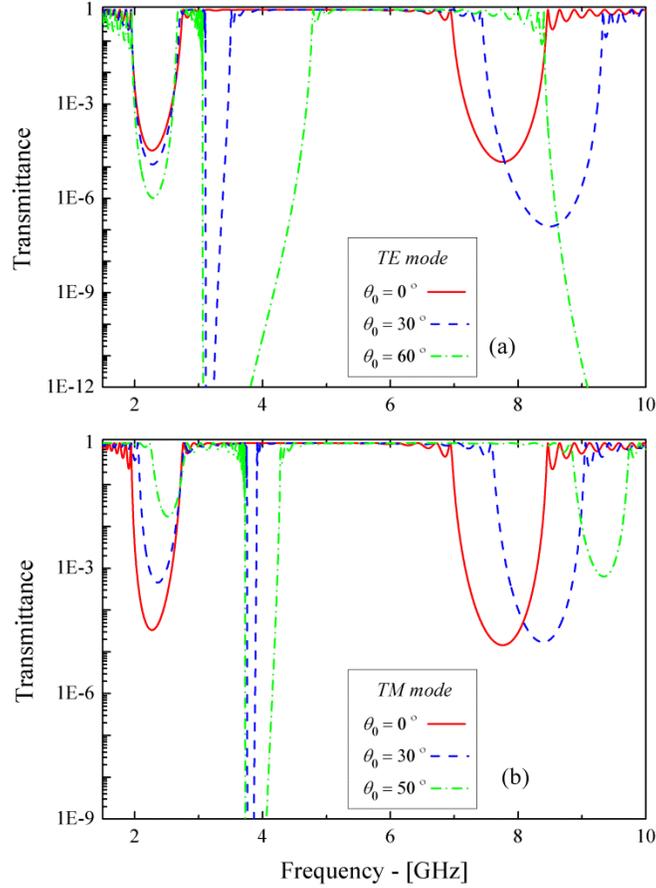

Fig. 2. Transmission spectra for the 1D MetaPC structure, for different angles of incidence, with the magnetic and electric loss factors are absent. i.e., $\gamma_e = \gamma_m = 0$, (a) TE mode, (b) TM mode.

In the first part, our study was focused on the normal incidence angle and investigated the behavior of zero-$\varepsilon$ and zero-$\mu$ gaps when we considered loss factors $\gamma_e$ and $\gamma_m$ for DNG material. Fig. 3 demonstrates the transmission spectra structure for normal incidence case and for different electric and magnetic loss factors. We observe that by applying $\gamma_e$ and $\gamma_m$, the rate of the transmittance and the basic characteristics of the Bragg and the zero-$\bar{n}$ gaps are affected. As reported in [14], an increase in the loss factors leads to the increase of the width of the zero-$\bar{n}$ gap and a decrease in its depth, while the central frequency remains nearly invariant. The rate of transmittance is also much dependant on the loss factors and decreases as $\gamma_e$ and $\gamma_m$ increases. The width, the depth, and the central frequency of the Bragg gap, compared to the zero-$\bar{n}$ gap, are almost invariant against the change in the electric and magnetic losses. Moreover, the Fig. 3 interestingly shows that when both loss factors are simultaneously considered, zero-$\mu$ and zero-$\varepsilon$ gaps appear in the transmission spectra. It is worth mentioning that these results are in sharp contrast with what has already been reported in [20,21]. It is clear that the zero-$\mu$ and zero-$\varepsilon$ band gaps appear in the transmission spectra for normal incidence case, and these gaps depend respectively on the electric ($\gamma_e$) and magnetic ($\gamma_m$) loss factors.

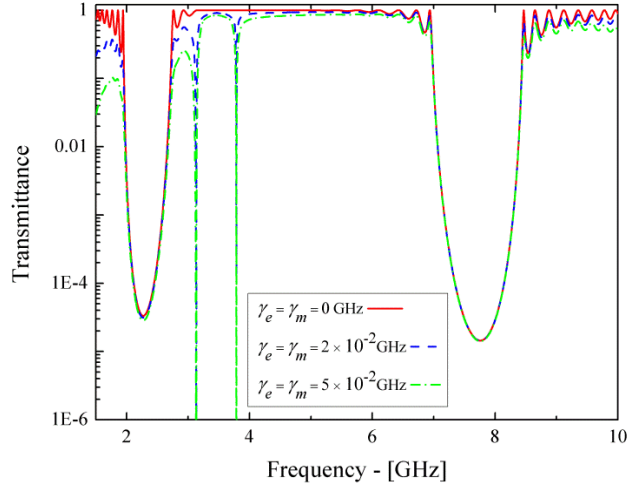

Fig. 3. Transmission spectra for the 1D MetaPC structure, for normal incidence case and for different magnetic and electric loss factors are absent.

In more detail, in Fig. 4, we have plotted the frequency-dependent transmittance for the normal incidence case at four different loss factors in the range of frequencies wherein the zero-$\mu$ and zero-$\varepsilon$ gaps appear when the electric and magnetic losses are equal, i.e., $\gamma_e = \gamma_m$. As it is observed in the figure, as loss factors increase, the increase in the width of the zero-$\mu$ gap is greater than that of the zero-$\varepsilon$ gap. In fact, to put it another way, the width of the zero-$\mu$ gap is more sensitive to the loss factors, compared to the other one. Additionally, the figure indicates that the rate of the transmittance is sensitive to the losses and it decreases as the loss factors increase. Besides, we observed that, the rate of the transmittance is more affected in the range of the frequency in which is the zero-$\mu$ gap appears.

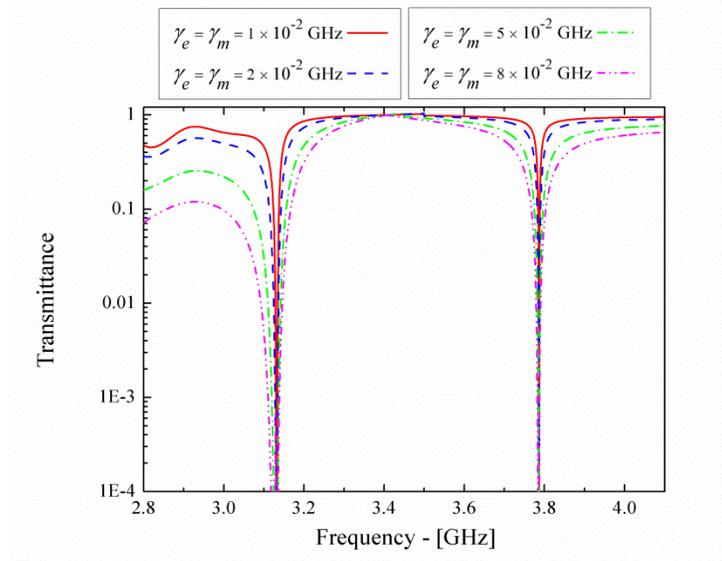

Fig. 4. Transmission spectra for the 1D MetaPC structure in the range of the frequency in which the zero-$\mu$ and zero-$\varepsilon$ gaps appear, for normal incidence case, with four different magnetic and electric loss factors when $\gamma_e = \gamma_m$.

Now, we separately investigate the role that the electric and magnetic loss factors play on zero-$\mu$ and zero-$\varepsilon$ gaps. Fig. 5 demonstrates the transmission spectra structure for normal incidence case and four different magnetic losses whereas the electric loss factor is neglected, i.e., $\gamma_e = 0$. It is clearly observed that, with increasing the $\gamma_m$, the width of the zero-$\mu$ gap increases, but the rate of the transmittance decreases. As expected, this increase in the bandwidth and the decrease in the transmittance are less than the situation in which both loss factors had been incorporated. In addition, as it is seen from the figure, it is interesting to note that when the $\gamma_e$ goes to zero, only the zero-$\mu$ gap continues to exist and the zero-$\varepsilon$ gap disappears.

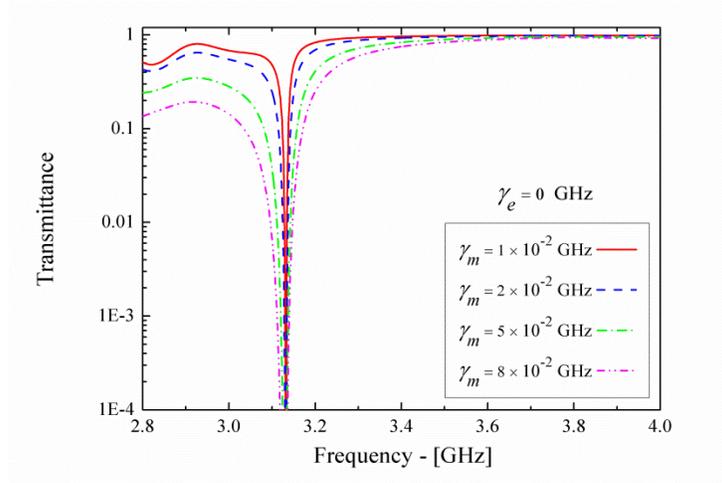

Fig. 5. Transmission spectra for the 1D MetaPC structure for normal incidence case, with four different magnetic loss factors when the electric loss factor is neglected.

In this part, we examine the role played by the electric loss $\gamma_e$ in the zero-$\varepsilon$ gap, wherein the magnetic loss factor $\gamma_m$ is taken as zero, i.e., $\gamma_m = 0$. In Fig. 6, we have plotted the transmission spectra for four different electric loss factors and at normal incidence case when the magnetic loss is neglected. As seen from the figure, the width of the zero-$\varepsilon$ gap increases and the transmittance decreases as the $\gamma_e$ increases. To compare with the results mentioned in the previous part in Fig. 5, we clearly see that these changes are less pronounced when the magnetic loss is incorporated and $\gamma_e = 0$. Moreover, it is also worth mentioning that another feature in Fig. 6 is that in the case of $\gamma_m = 0$, only the zero-$\varepsilon$ gap appears in the transmission spectra and the zero-$\mu$ gap vanishes.

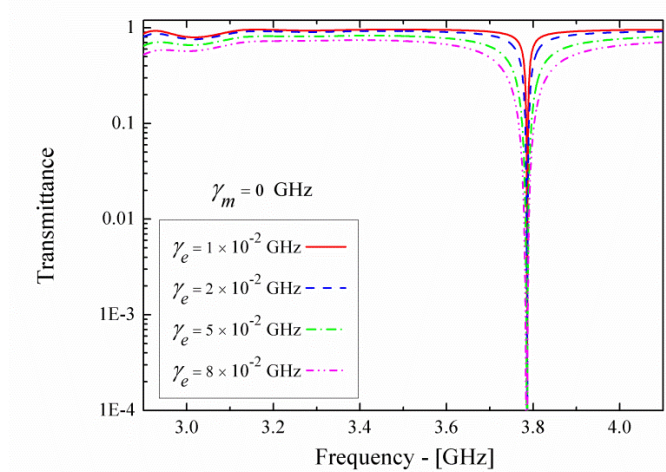

Fig. 6. Transmission spectra for the 1D MetaPC structure for normal incidence case, with four different electric loss factors when the magnetic loss factor is neglected.

In comparison with the previous situations in which only $\gamma_e$ or $\gamma_m$ were considered (Figs. 5 and 6), and also compared to Fig. 4 which displayed the simultaneous existence of the electric and magnetic loss factors, it is evident that the width of the zero-$\mu$ gap and the rate of the transmittance around the frequencies areas of this band gap are more affected by loss factors when compared to the zero-$\varepsilon$ gap. Additionally, it is expected that the changes in the bandwidth and the transmittance of both the zero-$\varepsilon$ and the zero-$\mu$ gaps in which $\gamma_e$ and $\gamma_m$ are simultaneously considered be more than the situation in which only one of the electric or magnetic losses are incorporated.

In the final part, we focused on the oblique incidence angle and investigated the behavior of both zero-$\varepsilon$ and zero-$\mu$ gaps when we considered loss factors $\gamma_e$ and $\gamma_m$ for DNG material. In Fig. 7, we have plotted the frequency-dependent transmittance in TE polarized wave for four different incidence angles $\theta_0 = 0°, 10°, 20°$, and $30°$, in the range of the frequencies in which the zero-$\mu$ and zero-$\varepsilon$ gaps appear when the electric and magnetic losses are equal, $\gamma_e = \gamma_m = 2 \times 10^{-2}$ GHz. As the figure shows and we mentioned before, for $\theta_0 = 0°$ both zero-$\varepsilon$ and zero-$\mu$ gaps appear in the transmission spectrum. Moreover, it is clearly observed that, with increasing the angle of incidence, as expected and mentioned in [20,21], the width of the zero-$\mu$ gap increases. In addition, the width of the zero-$\varepsilon$ gap increases as the angle of incidence increases, which is less than the increase in the zero-$\mu$ gap's width.

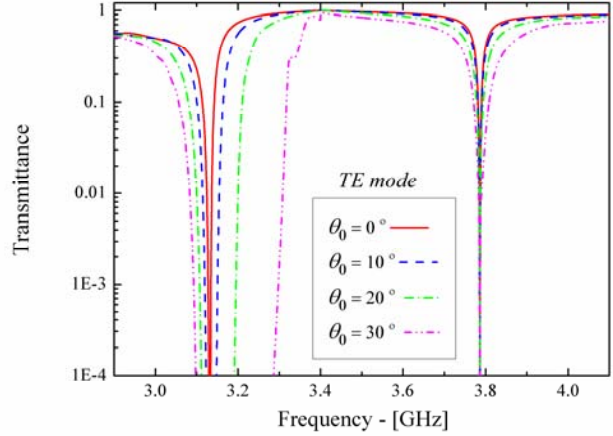

Fig. 7. Transmission spectra of TE polarized wave for the 1D MetaPC structure in the range of the frequencies in which the zero-$\mu$ and zero-$\varepsilon$ gaps appear, for four different angles of incidence, with $\gamma_e = \gamma_m = 2\times 10^{-2}$.

Fig. 8 demonstrates the transmission spectra structure for TM polarized wave and for four different incidence angles $\theta_0 = 0°, 10°, 20°$, and $30°$, whereas $\gamma_e = \gamma_m = 2\times 10^{-2}$ GHz, and similar to Fig. 7, in the range of the frequencies wherein the zero-$\mu$ and zero-$\varepsilon$ gaps appear. It is clearly observed that, with increasing the angle of incidence, the width of the zero-$\varepsilon$ gap increases. On the contrary, comparing with the zero-$\mu$ gap, the width of the zero-$\mu$ gap remains invariant when the angle of incidence increases.

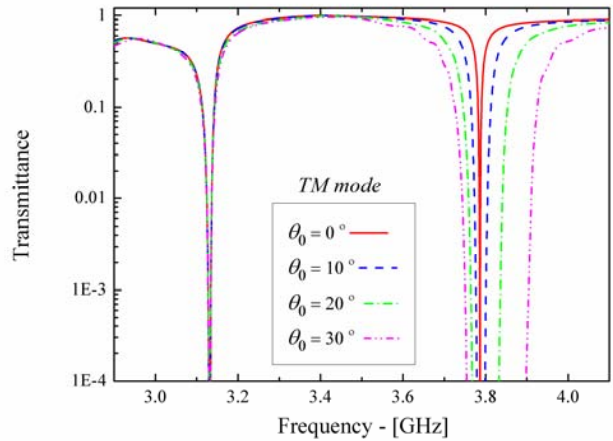

Fig. 8. Transmission spectra of TM polarized wave for the 1D MetaPC structure in the range of the frequencies in which the zero-$\mu$ and zero-$\varepsilon$ gaps appear, for four different angles of incidence, with $\gamma_e = \gamma_m = 2\times 10^{-2}$.

## 4. Conclusion

To summarize, this study has theoretically investigated the properties of the zero-$\mu$ and zero-$\varepsilon$ gaps in one-dimensional lossy photonic crystals composed of double-negative and double-positive materials. This study focuses on the impact of electric and magnetic loss factors, $\gamma_e$ and $\gamma_m$, on the zero-$\mu$ and zero-$\varepsilon$ gaps. Contrary to previous reports [20,21] which have mentioned that only for the oblique incidence angles in the transmission spectrum besides the Bragg gap and zero-$\bar{n}$ gap, zero-$\mu$ and zero-$\varepsilon$ gaps respectively appear for TE and TM polarized waves, and these two gaps do not exist at normal incidence case. Our numerical results show that by applying the electric and magnetic loss factors for double-negative materials, which are inevitable, both the zero-$\mu$ and zero-$\varepsilon$ gaps appear simultaneously between the conventional Bragg gap and zero-$\bar{n}$ gap in the transmission spectrum, and it is independent of the incidence angle and polarizations. In addition, and above all, the results of this study bring up the conclusion that the zero-$\mu$ and zero-$\varepsilon$ gaps appear not only for the oblique incidence but also for normal incidence case. Moreover, the results note that at normal incidence case, the appearance of the zero-$\mu$ and zero-$\varepsilon$ gaps is directly related to the magnetic and electric loss factors, respectively. And finally, the results show that with increasing the loss factors and the angle of incidence, the width of both gaps increases.

**Acknowledgements**

A. Aghajamali would like to acknowledge his gratitude to Parisa Shams for her help and useful discussion.